\documentstyle[preprint,aps]{revtex}

\newcommand{\bms}[1]{\mbox{\boldmath$#1$}}
\newcommand{\bmss}[1]{\mbox{\boldmath$\scriptstyle#1$}}
\newcommand{\tfrac}[2]{\mbox{$\frac{#1}{#2}$}}
\newcommand{\ul}{\underline}

\tightenlines

\begin{document}

\title{
\addtolength{\baselineskip}{-5pt}
Expressing the operations of quantum computing \\
in multiparticle geometric algebra
}
\draft
\author{ Shyamal S.\ Somaroo }
\address{
\addtolength{\baselineskip}{-5pt}
BCMP, Harvard Medical School \\
240 Longwood Ave., Boston, MA 02115
}
\author{ David G.\ Cory }
\address{
\addtolength{\baselineskip}{-5pt}
Dept.\ of Nuclear Engineering \\
Massachusetts Institute of Technology \\
Cambrdige, MA 02139
}
\author{ Timothy F.\ Havel\thanks{
To whom correspondence should be addressed
at {\tt havel@menelaus.med.harvard.edu}
(617/432-3242 office, 617/738-0516 FAX).}
} \address{
\addtolength{\baselineskip}{-5pt}
BCMP, Harvard Medical School \\
240 Longwood Ave., Boston, MA 02115 \\
\ \\
}
\date{\today}
\maketitle
\begin{abstract}
We show how the basic operations of quantum computing can be
expressed and manipulated in a clear and concise fashion using
a multiparticle version of geometric (aka Clifford) algebra.
This algebra encompasses the product operator formalism of
NMR spectroscopy, and hence its notation leads directly to
implementations of these operations via NMR pulse sequences.
\bigskip
\end{abstract}
\pacs{02.40.Dr,03.65.Fd,31.15.-p,33.25.+k}

Geometric (aka Clifford) algebra is a generalization
of vector algebra to arbitrary dimensions and signatures,
which provides a concise and geometrically transparent
notation for describing a wide range of physical
phenomena (for introductions and examples, see
\cite{HesteSobcz:84,Hest$NF1:86,Baylis:96,%
Lounesto:97,HavelNajfe:94,HavelNajfe:95a}).
Multiparticle geometric algebra is a recent further
generalization that enables one to deal with interacting
two state quantum systems \cite{DorLasGul:93,DoLaGuSoCh:96}.
Since most models of a quantum computer are based on
such systems \cite{Lloyd:93,Brassard:95,Divencenzo:95},
it is of interest to formulate the basic
operations of quantum computing in these terms,
with the goal of gaining deeper insight into them.
Given the recently discovered methods of
emulating a quantum computer via NMR spectroscopy
\cite{GershChuan:97,CorFahHav:97,CorPriHav:97},
it is further of interest to note that one of the
main tools that NMR spectroscopists have developed
to aid them in understanding their experiments,
known as the {\em product operator formalism\/}
\cite{SoEiLeBoEr:83,VenHilbers:83,ErnBodWok:87,BoulaRance:94a},
is a subalgebra of a multiparticle geometric algebra.
Geometric algebra also encompasses the quaternion
methods often used by NMR spectroscopists to calculate
the effects of composite pulses on product operators
\cite{BlumiSpies:85,CouLevErn:85,Siminovitch:97}.
Thus, as illustrated in this letter,
it is generally straightforward to translate
a quantum logic operation expressed
in the multiparticle geometric algebra
into an NMR implementation thereof.

We shall begin with a physically motivated definition
of the geometric algebra ${\cal G}(3)$ of a single particle.
This algebra is isomorphic to the Pauli matrix algebra,
where the isomorphism is obtained by identifying the
Pauli matrices $\ul{\bms\sigma}_{\,1}, \ul{\bms\sigma}_{\,2},
\ul{\bms\sigma}_{\,3}$ with an orthonormal basis $\bms\sigma_1,
\bms\sigma_2, \bms\sigma_3$ of a Euclidean vector space.
Note that the product of all three Pauli matrices
$\ul{\bms\sigma}_{\,1}\,\ul{\bms\sigma}_{\,2}\,\ul{\bms\sigma}_{\,3}$
is just the imaginary unit ``$\,\imath\,$'' times the identity matrix,
which we denote by $\ul{\bms\iota} \equiv \imath \ul{\bf 1}$.
This enables us to further identify the corresponding element
$\bms\iota \equiv \bms\sigma_1\,\bms\sigma_2\,\bms\sigma_3$ of
the geometric algebra ${\cal G}(3)$, which is commonly called
the {\em unit pseudo-scalar\/}, with the imaginary unit itself.
In this way ${\cal G}(3)$ becomes an algebra over
the real numbers, even though the Pauli matrix
algebra is usually regarded as a complex algebra.

The (nonrelativistic) $N$-particle geometric
algebra ${\cal G}^N(3)$ consists of a direct
product of $N$ copies of ${\cal G}(3)$:\footnote{
This definition can be derived from a relativistic
multiparticle theory based on the geometric algebra of
$N$ copies of space-time \cite{DorLasGul:93,DoLaGuSoCh:96}.}
\begin{equation}
{\cal G}^N(3) ~\equiv~ {\cal G}(3) \otimes \cdots
\otimes {\cal G}(3) \qquad \mbox{($N$ factors)}
\end{equation}
Throughout this letter, we shall assume that
$1 \le i,j,k \le 3$, $1 \le \ell,m,n \le N$,
and that $\alpha, \beta, \ldots$ are real scalars.
Then the multiparticle geometric algebra can also be
defined by the following five straightforward rules:
\begin{eqnarray}
\label{eq:rule1} &&
\{ \bms\sigma_i^\ell \}
\text{ is a basis for a real vector space;} \\
\label{eq:rule2} &&
\bms\sigma_i^\ell (\bms\sigma_j^m \bms\sigma_k^n) ~=~
(\bms\sigma_i^\ell \bms\sigma_j^m) \bms\sigma_k^n ~; \\
\label{eq:rule3} &&
\bms\sigma_i^\ell (\alpha \bms\sigma_j^m
+ \beta \bms\sigma_k^n) ~=~
\alpha \bms\sigma_i^\ell \bms\sigma_j^m +
\beta \bms\sigma_i^\ell \bms\sigma_k^n
\quad\mbox{and} \\
&& (\alpha \bms\sigma_i^\ell + \nonumber
\beta \bms\sigma_j^m) \bms\sigma_k^n ~=~
\alpha \bms\sigma_i^\ell \bms\sigma_k^n +
\beta \bms\sigma_j^m \bms\sigma_k^n ~; \\
\label{eq:rule4} &&
\bms\sigma_i^\ell \bms\sigma_j^\ell +
\bms\sigma_j^\ell \bms\sigma_i^\ell ~=~
2 \delta_{ij} \quad\text{(the Kronecker delta);} \\
\label{eq:rule5} &&
\text{for all $\ell \ne m$:} \quad
\bms\sigma_i^\ell \bms\sigma_j^m -
\bms\sigma_j^m \bms\sigma_i^\ell ~=~ 0 ~.
\end{eqnarray}
The NMR product operator formalism relies upon
a large repertoire of relatively complicated
rules for predicting the evolution of the spins,
which vary with the interaction (e.g. strong or weak coupling),
the total angular momentum (for spin $> \frac{1}{2}$), and so on.
With a little practice, however, all the rules
of the product operator formalism can be readily
derived as they are needed from the five rules
given in Eqs. (\ref{eq:rule1}--\ref{eq:rule5}).

By our definition of ${\cal G}(3)$ above,
a faithful matrix representation of this
$(2^3)$-dimensional real algebra is obtained
by mapping the basis vectors to the Pauli matrices:
$\bms\sigma_k \rightarrow \ul{\bms\sigma}_{\,k}$.
A matrix representation of ${\cal G}^N(3)$ is obtained simply
by taking $N$-fold Kronecker products of these matrices.
Since the representation by Pauli matrices is a {\em complex\/} algebra,
however, its {\em real\/} dimension is $2\cdot 4^N = 2^{2N+1}$,
which for $N>1$ is less than the dimension $2^{3N}$ of ${\cal G}^N(3)$.
It follows that the direct product representation is no longer faithful.
The extra degrees of freedom are due to the fact that,
unlike the matrix representation, there is a different
complex unit $\bms\iota^\ell \equiv \bms\sigma_1^\ell \,
\bms\sigma_2^\ell \, \bms\sigma_3^\ell$ for every particle.
Since the matrix algebra is believed to contain all
possible quantum states, operators and propagators,
these degrees of freedom have no known physical relevance.
They can easily be removed by multiplying
through by a primitive idempotent of the form
\begin{equation} \label{eq:cor}
{\bf C} ~\equiv~
\tfrac{1}{2} (1 - \bms\iota^1\bms\iota^2) \,
\tfrac{1}{2} (1 - \bms\iota^1\bms\iota^3) \, \cdots \,
\tfrac{1}{2} (1 - \bms\iota^1\bms\iota^N) ~.
\end{equation}
This {\em correlator\/} commutes
with everything in the product algebra,
and projects it onto an ideal of the correct dimension.
The projection can be interpreted physically as
locking the phases of the various particles together.
For this reason, we shall not specify which
factor our imaginary unit $\bms\iota$ comes from,
since all choices are rendered equal by the correlator
--- whose presence in all our expressions will also not,
in the interests of brevity, be written out explicitly.
Further discussion of these issues may be
found in \cite{DorLasGul:93,DoLaGuSoCh:96}.

Another class of primitive idempotents that we shall need are
\begin{equation}
{\bf E}_{\pm{}}^m ~\equiv~ \tfrac{1}{2} (1 \pm \bms\sigma_3^m) 
\end{equation}
and products thereof from different factors.
These are easily shown to have the following properties:
\begin{equation} \label{eq:idem_prop}
({\bf E}_{\pm{}}^m)^2 ~=~ {\bf E}_{\pm{}}^m ~, \quad
{\bf E}_{\pm{}}^m {\bf E}_{\mp{}}^m ~=~ 0 ~, \quad
{\bf E}_+^m + {\bf E}_-^m ~=~ 1 ~, \quad
\bms\sigma_3^m {\bf E}_{\pm{}}^m ~=~ \pm{\bf E}_{\pm{}}^m ~.
\end{equation}
A product of $N$ such idempotents, one for each particle,
has the $(2^N)\times(2^N)$ matrix representation
\begin{equation} \label{eq:idem_rep}
\ul{\bf E}_{\,\epsilon_1} \otimes \cdots \otimes \ul{\bf E}_{\,\epsilon_N}
~\leftrightarrow~
| \epsilon_1\cdots\epsilon_N \rangle\langle \epsilon_1\cdots\epsilon_N | ~,
\end{equation}
where $\epsilon_m = \pm 1$ and
$\ul{\bf E}_{\,\pm{}} \,\equiv\, \tfrac{1}{2}
(\ul{\bf 1} \,\pm\, \ul{\bms\sigma}_{\,3})$.
To illustrate the utility of these idempotents,
we will begin by using them to parametrize an
arbitrary entity from the multiparticle geometric
algebra ${\cal G}^N(3)$ in terms of ``rotors''
from its even subalgebra \cite{Hest$NF1:86}.
We shall do this for only two particles,
from which the general case should be clear.

Given an arbitrary multivector ${\bf M}
\in {\cal G}(3) \otimes {\cal G}(3)$,
we may write
\begin{equation}
{\bf M} ~=~ {\bf M} ({\bf E}_+^1 + {\bf E}_-^1) ({\bf E}_+^2 + {\bf E}_-^2)
~=~ {\bf M}{\bf E}_+^1{\bf E}_+^2 + \cdots + {\bf M}{\bf E}_-^1{\bf E}_-^2 ~.
\end{equation}
Let $\langle{\bf M}\rangle_+ \equiv \tfrac{1}{2} ({\bf M} + \hat{\bf M})$
denote the projection of ${\bf M}$ onto the even subalgebra
$({\cal G}(3) \otimes {\cal G}(3))^+ = {\cal G}^+(3) \otimes {\cal G}^+(3)$,
where $\hat{\bf M}$ is the grade involution of ${\bf M}$.\footnote{
For definitions of the grade involution, the reverse, and
other common geometric algebra terms, see \cite{Lounesto:97}.
The identity $({\cal G}(3) \otimes {\cal G}(3))^+ =
{\cal G}^+(3) \otimes {\cal G}^+(3)$ holds because
for all $1 \le i, j \le 3$ and $1 \le m, n \le N$,
$\bms\sigma_i^m\bms\sigma_j^n {\bf C} = -\bms\iota^m
\bms\sigma_i^m \bms\iota^n \bms\sigma_j^n {\bf C}$ where
${\bf C}$ is the correlator defined in Eq.\ (\ref{eq:cor}).}
Then, using the fact that ${\bf E}_{\pm{}}^m {\bf E}_{\mp{}}^m = 0$
and $\hat{\bf E}_{\pm{}}^m = {\bf E}_{\mp{}}^m$,
each term above may be written as
\begin{eqnarray}
{\bf M} {\bf E}_{\epsilon_1}^1 {\bf E}_{\epsilon_2}^2 ~
& =~ & ({\bf M} {\bf E}_{\epsilon_1}^1 {\bf E}_{\epsilon_2}^2)
({\bf E}_{\epsilon_1}^1 {\bf E}_{\epsilon_2}^2) \nonumber \\
& =~ & ({\bf M} {\bf E}_{\epsilon_1}^1 {\bf E}_{\epsilon_2}^2
+ \hat{\bf M} {\bf E}_{-\epsilon_1}^1 {\bf E}_{-\epsilon_2}^2)
({\bf E}_{\epsilon_1}^1 {\bf E}_{\epsilon_2}^2) \\
& =~ & 2 \langle{\bf M} {\bf E}_{\epsilon_1}^1 {\bf E}_{\epsilon_2}^2
\rangle_+ ({\bf E}_{\epsilon_1}^1 {\bf E}_{\epsilon_2}^2) \nonumber
\end{eqnarray}
($\epsilon_1,\epsilon_2 \in \{\pm 1\}$).
It follows that
\begin{equation}
{\bf M} ~=~
(\bms\psi_{++} {\bf E}_+^1 {\bf E}_+^2 +
\bms\psi_{+-} {\bf E}_+^1 {\bf E}_-^2 +
\bms\psi_{-+} {\bf E}_-^1 {\bf E}_+^2 +
\bms\psi_{--} {\bf E}_-^1 {\bf E}_-^2) ~,
\end{equation}
where each rotor $\bms\psi_{\epsilon_1\epsilon_2}$
is in ${\cal G}^+(3) \otimes {\cal G}^+(3)$,
and each $\bms\psi_{\epsilon_1\epsilon_2}
{\bf E}_{\epsilon_1}^1{\bf E}_{\epsilon_2}^2$
can be viewed as a two-particle spinor
\cite{DorLasGul:93,DoLaGuSoCh:96}.
A similar expansion holds for any number of particles and,
when translated into matrices via Eq.\ (\ref{eq:idem_rep}) above,
corresponds to an expansion of the matrix into a sum of matrices
each of which contains one column from the original matrix,
and is otherwise zero.

If we multiply ${\bf M}$ by its reverse $\tilde{\bf M}$, we obtain
\begin{eqnarray} \label{eq:dm}
{\bf M}\tilde{\bf M} ~&=~&
(\bms\psi_{++} {\bf E}_+^1 {\bf E}_+^2 +
\cdots + \bms\psi_{--} {\bf E}_-^1 {\bf E}_-^2)
({\bf E}_+^1 {\bf E}_+^2 \tilde{\bms\psi}_{++} +
\cdots + {\bf E}_-^1 {\bf E}_-^2 \tilde{\bms\psi}_{--}) \\ \nonumber
&=~& \bms\psi_{++} {\bf E}_+^1 {\bf E}_+^2 \tilde{\bms\psi}_{++} +
\bms\psi_{+-} {\bf E}_+^1 {\bf E}_-^2 \tilde{\bms\psi}_{+-} +
\bms\psi_{-+} {\bf E}_-^1 {\bf E}_+^2 \tilde{\bms\psi}_{-+} +
\bms\psi_{--} {\bf E}_-^1 {\bf E}_-^2 \tilde{\bms\psi}_{--} ~.
\end{eqnarray}
Since the reverse corresponds to the Hermitian
conjugate of matrices, this expression
represents a general Hermitian matrix.
A rank one Hermitian matrix is obtained
from any single term of the expression, e.g.
$\bms\psi_{++}{\bf E}_+^1{\bf E}_+^2\tilde{\bms\psi}_{++}$.
Provided that $\bms\psi_{++}\tilde{\bms\psi}_{++} = 1$,
the scalar part of this expression is $\frac{1}{4}$,
which corresponds to a matrix trace of one.
Thus this expression represents the density matrix
of a pure state, while a general density matrix is a
convex combination of terms as in Eq.\ (\ref{eq:dm}) above.

A simple rotation by an angle $\alpha$ about
an axis ${\bf a} = \Sigma_i a_i \bms\sigma_i$
($\|{\bf a}\| = 1$) in a one-particle space
is given by the exponential of its generator,
e.g.\ $\exp(-\alpha\bms\iota{\bf a}/2)$,
just as is commonly done with Pauli matrices.
Due to the commutivity of vectors $\bms\sigma_i^m$,
$\bms\sigma_j^n$ from different particle spaces,
one cannot express arbitrary rotations
{\em between\/} particle spaces in the algebra.
An exception to this rule are the
{\em particle interchange operators\/},
which in the case of two particles is given by
\begin{equation}
\bms\Pi^{1,2} ~\equiv~
\tfrac{1}{2} (1 + \bms\sigma_1^1 \bms\sigma_1^2 +
\bms\sigma_2^1 \bms\sigma_2^2 + \bms\sigma_3^1 \bms\sigma_3^2) ~.
\end{equation}
An important variation on this is given by the
{\em scalar coupling Hamiltonian\/} in NMR spectroscopy,
${\bf J}^{1,2} \equiv \pi J^{1,2} (2 \bms\Pi^{1,2} - 1) / 2$,
where $J^{1,2}$ is a coupling constant in Hertz \cite{ErnBodWok:87}.
Since it is easily shown that $\bms\Pi^{1,2}$ squares to one,
the corresponding time-dependent propagator is
\begin{equation}
\exp(\bms\iota {\bf J}^{1,2} t) ~=~
e^{-\bmss{\iota} \pi J^{1,2} t / 2}
(\cos(\pi J^{1,2} t) + \bms\iota \bms\Pi^{1,2} \sin(\pi J^{1,2} t)) ~.
\end{equation}
In the product operator formalism, one generally assumes
{\em weak coupling\/}, i.e.\ $| \omega^1 - \omega^2 | \gg \pi J^{1,2}$,
where $\omega^1,\omega^2$ are the resonance frequencies of the spins.
This enables the usual two-spin Hamiltonian of NMR spectroscopy,
i.e.\ ${\bf H}^{1,2} ~\equiv~ {\bf J}^{1,2} + {\bf K}^{1,2}$
where ${\bf K}^{1,2} ~\equiv~ \frac{1}{2}(\omega^1\bms\sigma_3^1
+ \omega^2\bms\sigma_3^2)$ denotes the {\em Zeeman Hamiltonian\/},
to be replaced by its first-order approximation
\begin{equation}
{\bf H}_{\rm weak}^{1,2} ~\equiv~ {\bf K}^{1,2} +
\tfrac{1}{2} \pi J^{1,2} \bms\sigma_3^1 \bms\sigma_3^2 ~.
\end{equation}
Since this approximation is diagonal in the
usual $\bms\sigma_3^1 + \bms\sigma_3^2$ basis,
its propagator can be written down in closed form
(see e.g.\ \cite{SoEiLeBoEr:83,VenHilbers:83,ErnBodWok:87}).
Geometric algebra is however {\em not\/} limited
to weak coupling, although strong coupling of
course complicates the propagators substantially.
This will be demonstrated shortly, when we show
how to diagonalize a general two-spin Hamiltonian.

We next show how to implement a rotation in one particle space
{\em conditional\/} on the state of another it is correlated with.
The simplest example is the controlled-NOT (or XOR) quantum gate,
which is generated by a {\em transition Hamiltonian\/} of the form
\begin{equation}
{\bf H}_{\rm tr}^{1|2} ~\equiv~ \bms\sigma_1^1 {\bf E}_-^2
~=~ \tfrac{1}{2} \bms\sigma_1^1 (1 - \bms\sigma_3^2) ~.
\end{equation}
Since $(\bms\sigma_1^1)^2 = 1$ and $\bms\sigma_1^1$
commutes with the idempotent ${\bf E}_-^2$,
the corresponding propagator is given by
\begin{equation}
{\bf R}_1^{1|2}(\alpha) ~=~
e^{-\bmss{\iota} \alpha {\bf H}_{\rm tr}^{1|2} /2} ~=~
e^{-\bmss{\iota} \alpha \bmss{\sigma}_1^1/2}
{\bf E}_-^2 + {\bf E}_+^2 ~.
\end{equation}
This rotates the first spin about $\bms\sigma_1^1$ in those states
in which the second spin is along the $-\bms\sigma_2^3$ axis.
The controlled-NOT ${\bf R}_1^{1|2} \equiv {\bf R}_1^{1|2}(\pi)$
is obtained when $\alpha = \pi$,
in which case the density matrix of the basis states
${\bf E}_{\epsilon_1}^1 {\bf E}_{\epsilon_2}^2$
are transformed as follows:
\begin{equation}
{\bf R}_1^{1|2} ({\bf E}_{\epsilon_1}^1
{\bf E}_{\epsilon_2}^2) \tilde{\bf R}_1^{1|2}
~=~ {\bf E}_{\epsilon_1\epsilon_2}^1 {\bf E}_{\epsilon_2}^2
~=~ \left\{ \begin{array}{ll}
{\bf E}_{-\epsilon_1}^1 {\bf E}_{\epsilon_2}^2
& \mbox{if $\epsilon_2 = -1$} \\
{\bf E}_{+\epsilon_1}^1 {\bf E}_{\epsilon_2}^2
& \mbox{if $\epsilon_2 = +1$}
\end{array} \right.
\end{equation}
In order to obtain a controlled-NOT that preserves
the phases of the basis states (as is usually assumed),
one need only multiply ${\bf R}_1^{1|2}$ by the conditional
phase shift $\exp(\bms\iota\pi{\bf E}_-^2/2)
= \bms\iota {\bf E}_-^2 + {\bf E}_+^2$.
Similarly, the propagator for the controlled-controlled-NOT
(or Toffoli gate) can be written as
\begin{equation}
e^{\bmss\iota\pi (1 - \bmss\sigma_1^1) {\bf E}_-^2 {\bf E}_-^3 / 2} ~=~
\bms\sigma_1^1 {\bf E}_-^2 {\bf E}_-^3 + (1 - {\bf E}_-^2 {\bf E}_-^3) ~,
\end{equation}
while the Fredkin gate is given by the conditional particle interchange
\begin{equation}
e^{\bmss\iota\pi (1 - \bmss\Pi^{1,2}) {\bf E}_+^3 / 2} ~=~
\bms\Pi^{1,2} {\bf E}_+^3 + {\bf E}_-^3 ~.
\end{equation}

In \cite{CorPriHav:97}, implementations of the controlled-NOT
and Toffoli gates by NMR pulse sequences were given,
which can also be easily validated by geometric algebra methods.
In the case of the controlled-NOT, the implementation consists
of a product of three propagators, which may be simplified as
\begin{eqnarray}
{\bf S}_1^{1|2} &~\equiv~& e^{-\bmss\iota\pi\bmss\sigma_1^1/4}
e^{-\bmss\iota\pi\bmss\sigma_3^1\bmss\sigma_3^2/4}
e^{-\bmss\iota\pi\bmss\sigma_2^1/4} \nonumber \\
&~=~& e^{-\bmss\iota\pi\bmss\sigma_1^1/4}
(e^{-\bmss\iota\pi\bmss\sigma_3^1\bmss\sigma_3^2/4}
e^{-\bmss\iota\pi\bmss\sigma_2^1/4} 
e^{\bmss\iota\pi\bmss\sigma_3^1\bmss\sigma_3^2/4})
e^{-\bmss\iota\pi\bmss\sigma_3^1\bmss\sigma_3^2/4} \\
&~=~& e^{-\bmss\iota\pi\bmss\sigma_1^1/4}
e^{\bmss\iota\pi\bmss\sigma_1^1\bmss\sigma_3^2/4}
e^{-\bmss\iota\pi\bmss\sigma_3^1\bmss\sigma_3^2/4}
({\bf E}_+^2 + {\bf E}_-^2) \nonumber \\
&~=~& (e^{-\bmss\iota\pi\bmss\sigma_1^1/2} {\bf E}_-^2 + {\bf E}_+^2)
(e^{-\bmss\iota\pi\bmss\sigma_3^1/4} {\bf E}_+^2 +
e^{\bmss\iota\pi\bmss\sigma_3^1/4} {\bf E}_-^2) \nonumber \\
&~=~& e^{-\bmss\iota\pi\bmss\sigma_1^1/2}
e^{\bmss\iota\pi\bmss\sigma_3^1/4} {\bf E}_-^2 +
e^{-\bmss\iota\pi\bmss\sigma_3^1/4} {\bf E}_+^2 ~, \nonumber
\end{eqnarray}
where we have used Eq.\ (\ref{eq:idem_prop}) to get the relation
$\exp(-\bms\iota\pi\bms\sigma_3^1\bms\sigma_3^2/4) {\bf E}_{\pm{}}^2
= \exp(\mp\bms\iota\pi\bms\sigma_3^1/4) {\bf E}_{\pm{}}^2$.
To get a controlled-NOT that preserves phases,
we may left-multiply ${\bf S}_1^{1|2}$ by
$\exp(-\bms\iota\pi(1+\bms\sigma_3^1-\bms\sigma_3^2)/4)$
as in \cite{CorPriHav:97}, or right-multiply it
by $\exp(-\bms\iota\pi\bms\sigma_3^1{\bf E}_-^2/4)
\exp(\bms\iota\pi\bms\sigma_3^1{\bf E}_+^2/4) =
\exp(\bms\iota\pi\bms\sigma_3^1\bms\sigma_3^2/4)$
along with $\exp(\bms\iota\pi{\bf E}_-^2/2)$ as above.
The validation of the Toffoli pulse sequence
is similar, though considerably more complex.

An interesting generalization of the two particle
conditional rotations is obtained by taking
the exponential of all four transitions, i.e.
\begin{eqnarray}
\nonumber && \exp( -\bms\iota (
\alpha_+ \bms\sigma_1^1 {\bf E}_+^2 +
\alpha_- \bms\sigma_1^1 {\bf E}_-^2 +
\beta_+ \bms\sigma_1^2 {\bf E}_+^1 +
\beta_- \bms\sigma_1^2 {\bf E}_-^1 ) / 2 ) \\
&=~& \exp( -\bms\iota ( (\alpha_+ + \alpha_-) \bms\sigma_1^1 / 4
+ (\beta_+ - \beta_-) \bms\sigma_3^1 \bms\sigma_1^2 / 4 ) ) \\
\nonumber && \exp( -\bms\iota ( (\alpha_+ - \alpha_-) \bms\sigma_1^1
\bms\sigma_3^2 / 4 + (\beta_+ + \beta_-) \bms\sigma_1^2 / 4 ) ) \\
&\equiv~& \exp( -\bms\iota {\bf X} ) \exp( -\bms\iota {\bf Y} ) ~=~
\exp( -\bms\iota {\bf Y} ) \exp( -\bms\iota {\bf X} ) \nonumber ~,
\end{eqnarray}
where ${\bf X}$ and ${\bf Y}$ commute.
In addition, since
\begin{eqnarray}
{\bf X}^2 &~=~& ((\alpha_+ + \alpha_-)^2 + (\beta_+ - \beta_-)^2) / 16 \\
{\bf Y}^2 &~=~& ((\alpha_+ - \alpha_-)^2 + (\beta_+ + \beta_-)^2) / 16 ~,
\nonumber
\end{eqnarray}
their propagators can be written in closed form as
\begin{eqnarray}
\exp( -\bms\iota {\bf X} ) &~=~& \cos(\sqrt{{\bf X}^2}) -
\bms\iota \sin(\sqrt{{\bf X}^2}) {\bf X} / \sqrt{{\bf X}^2} \\
\exp( -\bms\iota {\bf Y} ) &~=~& \cos(\sqrt{{\bf Y}^2}) -
\bms\iota \sin(\sqrt{{\bf Y}^2}) {\bf Y} / \sqrt{{\bf Y}^2} ~.
\nonumber
\end{eqnarray}

For example, if $\alpha_+ = \beta_+ = 0$ and
$\alpha_- = \beta_- = \pi\sqrt{2}$, we obtain
$(\bms\sigma_3^1 + \bms\sigma_3^2 -
\bms\sigma_1^1\bms\sigma_1^2 -
\bms\sigma_2^1\bms\sigma_2^2)/2$,
which is the particle interchange operator
$\bms\Pi^{1,2}$ up to a conditional phase of
$\exp(\bms\iota\pi{\bf E}_+^1{\bf E}_+^2)$.
If, on the other hand, $\alpha_+ = \beta_- = \pi \sqrt{2}$,
we get $(\bms\sigma_3^1 - \bms\sigma_3^2 - \bms\sigma_1^1\bms\sigma_1^2
+ \bms\sigma_2^1\bms\sigma_2^2)/2$, which is converted by
$\exp( \bms\iota\pi {\bf E}_+^1 {\bf E}_-^2 )$ into
$\frac{1}{2}(1 - \bms\sigma_3^1\bms\sigma_3^2 +
\bms\sigma_1^1\bms\sigma_1^2 - \bms\sigma_2^1\bms\sigma_2^2)$.
The terms $1 - \bms\sigma_3^1\bms\sigma_3^2$ are
annihilated by ${\bf E}_{\pm{}}^1 {\bf E}_{\pm{}}^2$,
while $\bms\sigma_1^1\bms\sigma_1^2 - \bms\sigma_2^1\bms\sigma_2^2$
induces direct transitions between the ${\bf E}_+^1{\bf E}_+^2$
and ${\bf E}_-^1{\bf E}_-^2$ states:
\begin{eqnarray}
&& (\bms\sigma_1^1\bms\sigma_1^2 - \bms\sigma_2^1\bms\sigma_2^2)
{\bf E}_{\pm{}}^1 {\bf E}_{\pm{}}^2
(\bms\sigma_1^1\bms\sigma_1^2 - \bms\sigma_2^1\bms\sigma_2^2) \\
&~=~& {\bf E}_{\mp{}}^1 {\bf E}_{\mp{}}^2
(\bms\sigma_1^1\bms\sigma_1^2 - \bms\sigma_2^1\bms\sigma_2^2)^2
~=~ {\bf E}_{\mp{}}^1 {\bf E}_{\mp{}}^2
(2 + 2 \bms\sigma_3^1\bms\sigma_3^2)
~=~ 4\, {\bf E}_{\mp{}}^1 {\bf E}_{\mp{}}^2 ~. \nonumber
\end{eqnarray}
We note that these transformations can be implemented
in NMR via what might be called {\em compound pulses\/},
i.e.\ multiple simultaneous pulses each selective for
a single transition \cite{SoEiLeBoEr:83,HatanYanno:81}.

Conditional rotations further enable us to diagonalize a
general two-spin NMR Hamiltonian ${\bf H}^{1,2}$ as above.
If we define the correlated idempotents ${\bf E}_{\pm{}}^{1,2}
\equiv \frac{1}{2}( 1 \pm \bms\sigma_3^1 \bms\sigma_3^2 )$,
this may be rewritten as
\begin{equation}
{\bf H}^{1,2} ~=~
\tfrac{1}{2} \omega^1 \bms\sigma_3^1 +
\tfrac{1}{2} \omega^2 \bms\sigma_3^2 +
\pi J^{1,2} \left( \tfrac{1}{2} \bms\sigma_3^1 \bms\sigma_3^2 +
\bms\sigma_1^1 \bms\sigma_1^2 {\bf E}_-^{1,2} \right) ~.
\end{equation}
We can split ${\bf H}^{1,2}$ into two commuting parts as follows:
\begin{equation}
{\bf H}^{1,2} ~=~ {\bf H}^{1,2} {\bf E}_+^{1,2} + {\bf H}^{1,2}
{\bf E}_-^{1,2} ~\equiv~ {\bf H}_+^{1,2} + {\bf H}_-^{1,2} ~.
\end{equation}
Since ${\bf E}_+^{1,2} {\bf E}_-^{1,2} = 0$,
the ${\bf H}_+^{1,2}$ part is already diagonal.
The off-diagonal part, on the other hand,
can be written as
\begin{equation}
{\bf H}_-^{1,2} - \tfrac{1}{2} \pi J^{1,2} {\bf E}_-^{1,2}
~=~ ( \omega_- \bms\sigma_3^1 + \pi J^{1,2}
\bms\sigma_1^1 \bms\sigma_1^2 ) {\bf E}_-^{1,2} ~,
\end{equation}
where $\omega_{\pm{}} \equiv \omega^1 \pm{} \omega^2$.
The $\bms\sigma_3^1$ and $\bms\sigma_1^1 \bms\sigma_1^2$
terms are rotated into one another by
$\exp(\phi \bms\iota \bms\sigma_2^1 \bms\sigma_1^2 )$,
and hence this can be written in polar form as
\begin{equation}
{\bf H}_-^{1,2} - \tfrac{1}{2} \pi J^{1,2} {\bf E}_-^{1,2}
~=~ \Theta \bms\sigma_3^1 \exp( \phi \bms\iota
\bms\sigma_1^1 \bms\sigma_1^2 ) {\bf E}_-^{1,2} ~,
\end{equation}
where $\Theta \equiv \sqrt{ (\omega_-)^2 + (\pi J^{1,2})^2 }$
and $\phi \equiv \arctan( \pi J^{1,2} / \omega_- )$.
The right-hand side of this equation is clearly diagonalized
by $\exp( -\phi \bms\iota \bms\sigma_2^1 \bms\sigma_1^2 / 2)$
and commutes with ${\bf E}_-^{1,2}$, from which it follows that
${\bf H}^{1,2}$ is diagonalized by the conditional rotation
\begin{equation}
{\bf T}(\phi) ~\equiv~ e^{-\phi \bmss\iota
\bmss\sigma_2^1 \bmss\sigma_1^2 {\bf E}_-^{1,2} / 2} ~=~
e^{-\phi \bmss\iota \bmss\sigma_1^1 \bmss\sigma_1^2 / 2}
{\bf E}_-^{1,2} + {\bf E}_+^{1,2} ~.
\end{equation}
The result is a weak coupling Hamiltonian whose
frequencies have been shifted by $\pm\Theta$,
\begin{equation}
\tilde{\bf T}(\phi) {\bf H}^{1,2} {\bf T}(\phi) ~=~
\tfrac{1}{2}(\omega_+ + \Theta) \bms\sigma_3^1 +
\tfrac{1}{2}(\omega_+ - \Theta) \bms\sigma_3^2 +
\tfrac{1}{2} \pi J^{1,2} \bms\sigma_3^1 \bms\sigma_3^2 ~,
\end{equation}
while the corresponding transition moment shows
that the peak intensities have been altered by
\begin{eqnarray}
&& \left\langle [ {\bf E}_{\epsilon_1}^1 
\tilde{\bf T}(\phi) (\bms\sigma_1^1 + \bms\sigma_1^2) {\bf T}(\phi)
{\bf E}_{\epsilon_1'}^1 ] [ {\bf E}_{\epsilon_2}^2
\tilde{\bf T}(\phi) (\bms\sigma_1^1 + \bms\sigma_1^2) {\bf T}(\phi)
{\bf E}_{\epsilon_2'}^2 ] \right\rangle \\
&=~& (1 + \epsilon_1 \sin(\phi))
\delta_{\epsilon_1,\epsilon_1'} \delta_{\epsilon_2,-\epsilon_2'}
+ (1 - \epsilon_2 \sin(\phi)) \nonumber
\delta_{\epsilon_1,-\epsilon_1'} \delta_{\epsilon_2,\epsilon_2'} ~.
\end{eqnarray}

The Hadamard transform plays an
essential role in many quantum algorithms,
but its simple geometric interpretation is seldom pointed out.
Consider a rotation of the $m$-th particle by an angle $\alpha$ about
the ${\bf w}^m \equiv (\bms\sigma_1^m + \bms\sigma_3^m)/\sqrt{2}$ axis:
\begin{equation}
{\bf W}^m(\alpha) ~\equiv~
\exp( -\bms\iota \alpha {\bf w}^m /2) ~=~
\cos(\alpha/2) - \bms\iota {\bf w}^m \sin(\alpha/2)
\end{equation}
The one-particle Hadamard transform ${\bf W}^m =
-\bms\iota{\bf w}^m$ is obtained when $\alpha = \pi$,
and the $N$-particle Hadamard transform
${\bf W}_N$ is simply the commutative product of
the Hadamard transforms of the individual particles.
The quantum Fourier transform (QFT) \cite{Shor$FAC:97}
can likewise be written, and simplified,
using the multiparticle geometric algebra.
As shown by Don Coppersmith (unpublished manuscript),
the $N$-particle QFT ${\bf Q}_N$ can be written
as a recursive product of one-particle Hadamard
transforms and two-particle conditional phase shifts,
\begin{equation}
{\bf V}^{\ell,m} ~\equiv~ \exp( -\bms\iota \omega_{\ell m}
{\bf E}_+^\ell {\bf E}_+^m ) \qquad (\omega_{\ell m}
\equiv \pi 2^{\ell - m} ~\mbox{for}~ \ell \le m) ~,
\end{equation}
namely
\begin{equation}
{\bf Q}_N ~\equiv~ {\bf U}^1 \cdots {\bf U}^{N-1} {\bf U}^N
~\equiv~ ({\bf W}^1 {\bf V}^{1,2} \cdots {\bf V}^{1,N})
\cdots ({\bf W}^{N-1} {\bf V}^{N-1,N}) ({\bf W}^N) ~.
\end{equation}

Each factor ${\bf U}^m$ above can be rearranged so
that all the Hadamaard transformations come first:
\begin{eqnarray}
{\bf U}^m &~=~& (-\bms\iota{\bf w}^m) \left(
e^{ -\bmss\iota \omega_{m,m+1} {\bf E}_+^m {\bf E}_+^{m+1} } \cdots
e^{ -\bmss\iota \omega_{m,N} {\bf E}_+^m {\bf E}_+^N } \right)
(\bms\iota{\bf w}^m) (-\bms\iota{\bf w}^m) \nonumber \\
&~=~& e^{ -\bmss\iota (-\bmss\iota{\bf w}^m) {\bf E}_+^m
(\bmss\iota{\bf w}^m) (\omega_{m,m+1} {\bf E}_+^{m+1}
+ \cdots + \omega_{m,N} {\bf E}_+^N ) } (-\bms\iota{\bf w}^m) \\
&~=~& e^{-\bmss\iota/2} e^{ -\bmss\iota \bmss\sigma_1^m
(\omega_{m,m+1} {\bf E}_+^{m+1} + \cdots + \omega_{m,N}
{\bf E}_+^N ) } (-\bms\iota{\bf w}^m) \nonumber
\end{eqnarray}
Since the factors ${\bf U}^n$ with $n > m$
do not involve the $m$-th particle space,
we may therefore write the QFT with all the
$-\bms\iota{\bf w}^m$ together on one side, i.e.
\begin{equation}
{\bf Q}_N ~=~ e^{-N\bmss\iota/2} e^{ -\bmss\iota \bmss\sigma_1^1
(\omega_{1,2} {\bf E}_+^2 + \cdots + \omega_{1,N} {\bf E}_+^N) } \cdots
e^{ -\bmss\iota \bmss\sigma_1^{N-1} (\omega_{N-1,N} {\bf E}_+^N) } {\bf W}_N ~.
\end{equation}
In NMR spectroscopy, each of the conditional
rotations in this expression can in principle
be implemented with a single compound pulse.
The Hadamard transform ${\bf W}_N$ of all $N$ spins,
on the other hand, can be implemented with three ``hard''
(nonselective) pulses, namely $\exp(\bms\iota\pi\bms\sigma_2/8)
\exp(\bms\iota\pi\bms\sigma_1/2) \exp(-\bms\iota\pi\bms\sigma_2/8)$,
which takes only a very small fraction of the time required
for a ``soft'' pulse (or the conditional rotations).
The complexity of this NMR implementation of the QFT is only $O(N)$,
and since the time required for a hard pulse is essentially
{\em independent\/} of the number of spins involved,
the implementation requires only half the number of steps
in the parallel implementation proposed by Coppersmith.

In closing, we note one final advantage
of multiparticle geometric algebra,
which is the astonishing ease with
which one can compute partial traces.
In terms of idempotents, the partial trace
of a density matrix $\bms\rho$ over any
one particle $m$ can be written as:
\begin{equation} \label{eq:ptr}
{\rm Tr}_m(\bms\rho) ~=~
{\bf E}_+^m \bms\rho\, {\bf E}_+^m +
{\bf E}_-^m \bms\rho\, {\bf E}_-^m +
\bms\sigma_1^m ({\bf E}_+^m \bms\rho\, {\bf E}_+^m +
{\bf E}_-^m \bms\rho\, {\bf E}_-^m) \bms\sigma_1^m ~.
\end{equation}
Consider, for example, the three particle GHZ state,
$\bms\rho_{\rm GHZ} ~\equiv$
\begin{equation}
(1 + \bms\sigma_3^1\bms\sigma_3^2
+ \bms\sigma_3^1 \bms\sigma_3^3 + \bms\sigma_3^2 \bms\sigma_3^3
+ \bms\sigma_1^1 \bms\sigma_1^2 \bms\sigma_1^3
- \bms\sigma_2^1 \bms\sigma_2^2 \bms\sigma_1^3
- \bms\sigma_2^1 \bms\sigma_1^2 \bms\sigma_2^3
- \bms\sigma_1^1 \bms\sigma_2^2 \bms\sigma_2^3) / 8 ~.
\end{equation}
Using the relations ${\bf E}_{\pm{}}^3 \bms\sigma_k^3 {\bf E}_{\pm{}}^3
= {\bf E}_{\pm{}}^3 {\bf E}_{\mp{}}^3 \bms\sigma_k^3 = 0$ and
$\bms\sigma_k^3{\bf E}_{\pm{}}^3\bms\sigma_k^3 = {\bf E}_{\mp{}}^3$
for $k = 1,2$, one finds that
\begin{equation}
{\bf E}_{\pm{}}^3 \bms\rho_{\rm GHZ} {\bf E}_{\pm{}}^3 ~=~
{\bf E}_{\pm{}}^3 (1 + \bms\sigma_3^1\bms\sigma_3^2
+ \bms\sigma_3^1 \bms\sigma_3^3 + \bms\sigma_3^2 \bms\sigma_3^3) / 8
\end{equation}
and
\begin{equation}
\bms\sigma_1^3 {\bf E}_{\pm{}}^3 \bms\rho_{\rm GHZ}
{\bf E}_{\pm{}}^3 \bms\sigma_1^3 ~=~
{\bf E}_{\mp{}}^3 (1 + \bms\sigma_3^1\bms\sigma_3^2
- \bms\sigma_3^1 \bms\sigma_3^3 - \bms\sigma_3^2 \bms\sigma_3^3) / 8 ~.
\end{equation}
The sum of the four terms in Eq.\ (\ref{eq:ptr})
is thus the mixed state
\begin{equation}
{\rm Tr}_3(\bms\rho_{\rm GHZ}) ~=~
(1 + \bms\sigma_3^1 \bms\sigma_3^2) / 4 ~.
\end{equation}
More generally, the partial trace is obtained simply by
dropping all terms from the density matrix containing
factors from the particles over which the trace is taken,
and multiplying by two to the number of such particles.

In conclusion, we have shown how to formulate
the most important operations of quantum
computing in multiparticle geometric algebra,
and illustrated some of the advantages that
this more general theory has over the established
product operator formalism used in NMR spectroscopy.
These advantages will be further demonstrated in subsequent
publications devoted to developing the theory and methods
needed for ensemble quantum computing by NMR spectroscopy.

\bigskip\centerline{\bf Acknowledgements}
This work was supported by, or in part by,
the U.\ S.\ Army Research Office under
contract/grant number DAAG 55-97-1-0342
from the DARPA Ultrascale Computing Program.

\bibliographystyle{plain}

\end{document}